\documentclass[prb,aps,showpacs,twocolumn,floatfix,10pt]{revtex4-1}

\pdfoutput=1

\usepackage[english]{babel}

\usepackage[utf8]{inputenc}

\usepackage{amsmath,amssymb,amstext}

\usepackage{txfonts}

\usepackage{lmodern}

\usepackage{xcolor,soul}

\usepackage{bm}

\usepackage{graphicx}

\usepackage{array}

\usepackage[caption=false]{subfig}

\usepackage[activate=normal]{pdfcprot}

\usepackage{siunitx}

\usepackage{multirow}

\usepackage{hyphenat}

\usepackage{marvosym}

\newlength{\graphicswidth}
\newlength{\graphicswidthfull}
\newcommand{\vect}[1]{\bm{#1}}

\newcommand*\diff{\mathop{}\!\mathrm{d}}

\newcommand*\pdiff{\mathop{}\!\mathrm{\partial}}

\DeclareMathOperator{\sign}{sign}

\definecolor{orange}{rgb}{1,0.5,0}
\definecolor{gray}{gray}{0.5}

\begin{document}

\title{Energy Conservation and Power Bonds in Co\hyp{}Simulations: Non-Iterative Adaptive Step Size Control and Error Estimation}
\author{Severin~Sadjina, Stian~Skjong, Eilif~Pedersen}
\affiliation{Department of Marine Technology, Norwegian University of Science and Technology, NO-7491 Trondheim, Norway}
\author{Lars~T.~Kyllingstad}
\affiliation{SINTEF Fisheries and Aquaculture, NO-7465 Trondheim, Norway}

\begin{abstract}
Here, we study the flow of energy between coupled simulators in a co\hyp{}simulation environment using the concept of power bonds.
We introduce energy residuals which are a direct expression of the coupling errors and hence the accuracy of co\hyp{}simulation results.
We propose a novel \emph{Energy\hyp{}Conservation\hyp{}based Co\hyp{}Simulation} method (ECCO) for adaptive macro step size control to improve accuracy and efficiency.
In contrast to most other co\hyp{}simulation algorithms, this method is non-iterative and only requires knowledge of the current coupling data.
Consequently, it allows for significant speed ups and the protection of sensitive information contained within simulator models.
A quarter car model with linear and nonlinear damping serves as a co\hyp{}simulation benchmark and verifies the capabilities of the energy residual concept:
Reductions in the errors of up to \SI{93}{\percent} are achieved at no additional computational cost.
\end{abstract}

\maketitle

\section{Introduction}
\label{sec:introduction}

Engineering systems continue to increase in complexity, comprising of all sorts of physical phenomena, and often characterized on vastly different time scales.
At the same time, development and manufacturing are constrained by an ever increasing pressure to keep costs low and time-to-market periods short.
Because of this, there is a strong virtualization trend in engineering.
Simulation methods play a crucial role in early product and system development where they allow to identify and avoid design flaws and risk potentials and replace time-intense and costly real testing.
The automotive and the maritime industries \cite{Henry2009,Harries2011,Palm2013,Bertsch2014} have recently, amongst others, increased their efforts to incorporate virtual prototyping into all stages of the development cycle.

Simulator coupling, or co\hyp{}simulation, has gained significant interest due to its attractiveness for scientific and industrial applications:
It allows for independent and parallel modeling, the efficient use and reuse of domain\hyp{}specific models, software tools, and expert knowledge, and significant simulation speed ups through parallelization and the use of suitable solvers within each simulator.\footnote{%
Note that we use the word `simulator' throughout in the sense of a \emph{subsimulator}: a mathematical model of a \emph{subsystem} coupled to other such models of \emph{subsystems} to form a full model of a total system.
}
Co\hyp{}simulation further enables the protection of intellectual property, which is often a strong requirement in industrial applications.
Its basic idea is the inclusion of a solver with each model in a loose coupling environment.

There are, however, two major challenges for this full-system simulation approach: stability and accuracy.
A general-purpose, easy-to-use and robust framework to estimate and control errors and ensure sufficient accuracy and stability is lacking.
Constant co\hyp{}simulation step sizes (macro step sizes) and simple synchronization algorithms can still be considered state-of-the-art.
In practice, the step size is often chosen manually by trial-and-error, sacrificing both accuracy and efficiency, and uncertainty estimates are simply not available.
Clearly, there is a need for more elaborate concepts that also provide easy-enough feedback about the quality of the co\hyp{}simulation results.
This is especially true for industrial applications:
The use of commercial software will in most cases prohibit iterative coupling schemes or at least make them computationally unfeasible.
Furthermore, the protection of simulator-internal data is typically a major concern and implementation details about the model or the internal solver method are unavailable.

Unfortunately, this excludes the use of almost all advanced co\hyp{}simulation schemes that have been proposed so far:
There exist a number of adaptive step size algorithms for co\hyp{}simulation \cite{Guenther2001,Verhoeven2008,Busch2011,Busch2012,Schierz2012,Arnold2013,Viel2014} which, based on a specific error estimation method, aim to choose an optimal macro step size to exchange coupling data efficiently.
However, practically all of these require simulator\hyp{}internal data (such as the time integration order) or the ability to revert to a previous time point and redo an entire macro time step (rollback).
Mostly, they require both.
One exception is the step size controller proposed in Refs.~\onlinecite{Busch2011,Busch2012} which adaptively controls the macro step size by use of a non\hyp{}iterative predictor/corrector error estimator.

A different approach is to correct for the coupling errors directly. In the \emph{Interface Jacobian\hyp{}based Co\hyp{}Simulation Algorithm} \cite{Sicklinger2014}, coupling conditions are solved iteratively with Newton's method.
While allowing for accurate solutions, it generally requires Jacobians along with the coupling variables from each simulator at each macro time step.
This, together with its iterative nature practically prohibits any industrial use.
An interesting non-iterative co\hyp{}simulation method is the \emph{Nearly Energy Preserving Coupling Element} \cite{Benedikt2013}.
It is based on the realization that the extrapolation of input signals itself can be considered an artificially introduced subsystem with its own error contributions.
These effectively change system behavior and violate the conservation of (generalized) energy.
Corrections to the coupling variables are then introduced in order to improve accuracy.

In the present paper, we take this approach in a different direction and directly study the flow of energy throughout a co\hyp{}simulation in detail using the concept of \emph{power bonds}.
We show that, if coupling variables are given in quantities whose product is a physical power---such as force and velocity, pressure and flow rate, or voltage and current---the energy exchanges between simulators are directly accessible.
Moreover, we investigate the concept of \emph{energy residuals} which directly alter the total energy of the overall coupled system due to the fact that the individual subsystems are solved independently of each other between communication points.
Successfully ensuring that all energy flows between simulators are (reasonably well) balanced is a strong indication for the validity of the co\hyp{}simulation results and a prerequisite to stability.
These findings are exploited in a novel \emph{Energy\hyp{}Conservation\hyp{}based Co\hyp{}Simulation} method (ECCO) which is generally applicable for simulator coupling due to its non-iterative nature and the fact that it is solely based on the current coupling data.

The paper is organized as follows:
In Section~\ref{sec:powerbonds}, we introduce the reader to power bonds and how they can be used to conveniently study the flow and conservation of energy in co\hyp{}simulations.
We also present the concept of power and energy residuals as a direct consequence of the inherent violation of the conservation of energy.
Next, we demonstrate their use for global error estimation to assess the quality of co\hyp{}simulation results in Section~\ref{sec:error_and_step_size}.
A quarter car co\hyp{}simulation benchmark model is used in Section~\ref{sec:benchmark_model} to demonstrate the performance of a non-iterative adaptive step size controller based on energy conservation.
The interesting effects of nonlinear damping and different system reticulations on accuracy and efficiency are also investigated for this benchmark model.
Section~\ref{sec:comparison_busch} provides a comparison to the predictor/corrector step size controller proposed by Busch \emph{et~al}.~\cite{Busch2011,Busch2012}.
Finally, a conclusion is given in Section~\ref{sec:conclusion}.

\section{Power Bonds and Energy Fluxes}
\label{sec:powerbonds}

Considering energy balances directly, that is, working with the Lagrangian or Hamiltonian equations is a poor starting point for co\hyp{}simulations because such ap\-proach\-es are based on the total energy of the complete system.
Instead, we shall follow the approach that is the theoretical foundation of bond graph theory \cite{Paynter1961} and describe the energy transactions between subsystems in terms of local energy continuity equations.\cite{Breedveld1984}
This way, we can balance the flow of energy for each subsystem separately and thereafter connect them all in a modular fashion while, in principle, satisfying the conservation of energy.

\subsection{Energy Continuity}		
\label{subsec:powerbonds_energycontinuity}

The transport of energy, that is, any energetic interaction between systems (or within systems) follows a general equation of continuity,
\begin{equation}
\label{eq:continuity_equation}
	\frac{\pdiff \epsilon(\vect{x},t)}{\pdiff t}
	+
	\vect{\nabla} \cdot \vect{j}_\epsilon(\vect{x},t)
	=
	\sigma_\epsilon(\vect{x},t)
	,
\end{equation}
where $\epsilon(\vect{x},t)$ is the local energy density at position $\vect{x}$ and time $t$, $\vect{j}_\epsilon$ is the energy flux, and $\sigma_\epsilon$ is the net rate of energy dissipation (or creation) per unit volume and unit time.
Eq.~\eqref{eq:continuity_equation} states that energy is locally continuous:
A change in the local energy density is always accounted for by appropriate local energy fluxes and energy dissipation (or creation).
With the modular structure of complex engineering systems and co\hyp{}simulation environments in mind, we are interested in already reticulated systems where the transport of energy between subsystems and the dissipation, creation, and distribution of energy inside of subsystems are typically restricted to discrete regions and strongly heterogeneous.
Under these assumptions, we can cast the energy continuity equation into a form much more suited for our purposes, \cite{Paynter1961}
\begin{equation}
\label{eq:continuity_equation_power}
	-\sum_\alpha P_\alpha(t)
	=
	\sum_\beta \frac{\diff E_\beta(t)}{\diff t}
	-
	\sum_\gamma \Sigma_\gamma(t)
	.
\end{equation}
Here, $P_\alpha$ is the total energy flux (that is, the power transmitted) through the \emph{power port} $k_\alpha$ of a subsystem, $E_\beta$ is the energy contained by a discrete sub-volume $V_\beta$ inside the subsystem, and, analogously, $\Sigma_\gamma$ is the rate of energy dissipation or creation inside a discrete sub-volume $V_\gamma$.
Eq.~\eqref{eq:continuity_equation_power} expresses the fact that any net flux of energy into a subsystem is either stored or dissipated.
In other words, if we correctly account for all net power transmitted between the subsystems as well as the energy stored and dissipated within the subsystems, we inevitably guarantee energy conservation and continuity.

\begin{figure}[h!tb]
	\centering
	\def\svgwidth{\graphicswidth}
	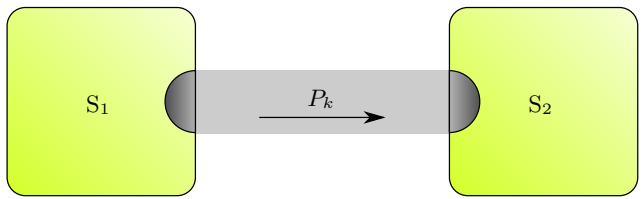
	\caption{%
		Two coupled subsystems, S$_1$ and S$_2$, exchange energy through a power bond $k$ at a rate $P_k = e_k f_k$ which is determined by the product of a flow $f_k$ and an effort $e_k$
	}
	\label{fig:power_bond}
\end{figure}

Naturally, we have little control over energy storage and dissipation and whether they are correctly accounted for inside simulators when they are coupled in a general co\hyp{}simulation environment.
We can, however, safely conclude that no energy should get lost,\footnote{%
This is true if we assume an ideal bond that does not include physical energy sources or sinks.
This is well justified in a co\hyp{}simulation setting where (relevant) energy creation and dissipation ought to be accounted for inside subsystems.
Our discussion is not limited by this choice, however, and it is fully possible to work with non-ideal bonds.
}
\begin{equation}
\label{eq:continuity_equation_powerbond}
	-
	(
	P_{k_1}
	+
	P_{k_2}
	)
	=
	0
	,
\end{equation}
while it is being transmitted \emph{between} the power ports $k_1$ and $k_2$ of two coupled simulators S$_1$ and S$_2$.
The energetic coupling between two subsystems via power ports is an example of a \emph{power bond}, see Fig.~\ref{fig:power_bond}.
Such a bond is defined by a pair of \emph{power variables}: a \emph{flow} and an \emph{effort}.
If we consider mechanical translation, for example, the flow $f_k$ is a velocity and the effort $e_k$ is a force.
The product of two power variables is always a power, $P_k = e_k f_k$, and gives the energy flow through the power bond (the rate at which energy is being exchanged).
This makes it practical to keep track of the flow and the conservation of energy and is the basis for bond graph theory.

\subsection{The Power Bond in a Co\hyp{}Simulation}
\label{subsec:powerbonds_cosimulation}

In a co\hyp{}simulation, the models include their own solvers that evolve their internal states $\vect{x}$ between the discrete communication time points $t_i$ and $t_{i+1}$,
\begin{subequations}
\label{equ:cosimulation}
\begin{equation}
\label{equ:cosimulation_evolution}
	\dot{\vect{x}}(t)
	=
	\vect{f}
	\big(
		\vect{x}(t)
		,
		\tilde{\vect{u}}(t)
	\big)
	,
	\quad
	t \in (t_i,t_{i+1}]
	,
\end{equation}
after having received inputs $\vect{u}$ from each other.
The values of the input variables are generally unknown between communication points, and an extrapolation $\tilde{\vect{u}}(t) \approx \vect{u}(t)$ has to be used.\footnote{%
Note that some coupling schemes \emph{interpolate} input values.
In the Gauss-Seidel iteration pattern, for example, one simulator is stepped first using input value extrapolation, and thereafter the second simulator is stepped using interpolation of the other system's outputs.
}
Often, the inputs are simply held constant during the time integration, hence $\tilde{\vect{u}}(t) = \vect{u}(t_i)$ for $t \in (t_i,t_{i+1}]$.
Outputs $\vect{y}$ are then computed from the internal states,
\begin{equation}
\label{equ:cosimulation_outputs}
	\vect{y}(t_{i+1})
	=
	\vect{g}
	\big(
		\vect{x}(t_{i+1})
		,
		\tilde{\vect{u}}(t_{i+1})
	\big)
	,
\end{equation}
and used as inputs again.\footnote{%
The functions $\vect{f}$ and $\vect{g}$ may also depend on time explicitly.
}
We can express the simulator coupling as
\begin{equation}
\label{equ:cosimulation_connections}
	\vect{u}(t_{i+1})
	=
	\vect{L}
	\vect{y}(t_{i+1})
	,
\end{equation}
\end{subequations}
where $\vect{L}$ is a connection graph matrix that relates outputs and inputs at communication time points.

\begin{figure}[h!tb]
	\centering
	\subfloat[Inputs are set at $t = t_i$]{%
		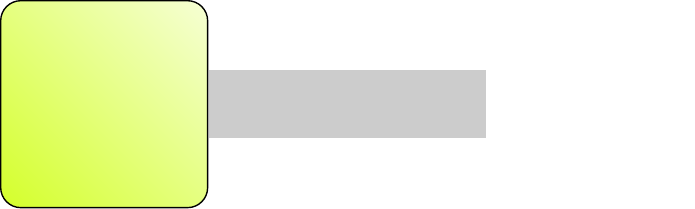
		\label{fig:power_bond_cosim_1}
	}\\
	\subfloat[Outputs are retrieved at $t = t_{i+1}$ after time integrations]{%
		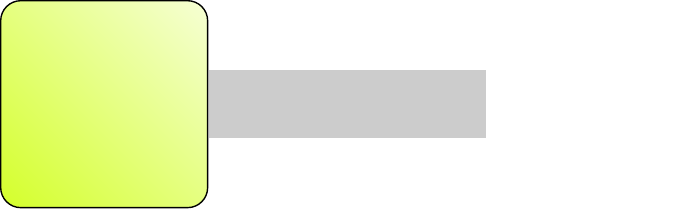
		\label{fig:power_bond_cosim_2}
	}
	\caption{%
		Two coupled simulators, S$_1$ and S$_2$, exchange energy through a power bond in a co\hyp{}simulation
		}
	\label{fig:power_bond_cosim}
\end{figure}

Now that we have set the stage, let us investigate the energy flow through a power bond $k$ between two simulators in a co\hyp{}simulation setting, see Fig.~\ref{fig:power_bond_cosim}.
Assuming that one variable (either the input or the output) represents a flow and the other an effort---which is the case for a power bond---this is easily done.
Simulator S$_1$ will conclude that the power it transmits to the other simulator S$_2$ through its power port $k_1$ is given by
\begin{subequations}
\label{equ:energyflow}
\begin{equation}
\label{equ:energyflow_S1}
	P_{k_1}(t)
	=
	\mbox{$\tilde{u}$}_{k_1}(t)
	y_{k_1}(t)
	,
\end{equation}
where $\mbox{$\tilde{u}$}_{k_1}$ and $y_{k_1}$ are the input and output, respectively.
Simulator S$_2$, on the other hand, will report that the power it received from S$_1$ through its power port $k_2$ is indeed
\begin{equation}
\label{equ:energyflow_S2}
	P_{k_2}(t)
	=
	\mbox{$\tilde{u}$}_{k_2}(t)
	y_{k_2}(t)
	,
\end{equation}
\end{subequations}
which leaves us with a predicament because generally $-(P_{k_1} + P_{k_2}) \neq 0$, in violation of Eq.~\eqref{eq:continuity_equation_powerbond}.

Before we have a closer look at this discrepancy and its consequences, let us first define an approximation for the total power transmitted through the power bond $k$ from simulator S$_1$ to simulator S$_2$ using both simulator outputs,
\begin{equation}
\label{equ:energytransfer}
	P_{k_{12}}(t)
	=
	\sigma_{k_{12}}
	\big(
		y_{k_1}(t)
		y_{k_2}(t)
	\big)
	.
\end{equation}
Here, the sign $\sigma_{k_{12}} \equiv ({L_k}_{12}-{L_k}_{21})/2$ is determined by the corresponding elements of the connection graph matrix $\vect{L}$.
This is very useful in studying the flow of energy throughout a co\hyp{}simulation, as we shall see in Sec.~\ref{sec:benchmark_model}.

\subsection{Residual Powers and Energies}
\label{subsec:powerbonds_residuals}

\begin{figure}[h!tb]
	\centering
	\def\svgwidth{\graphicswidth}
	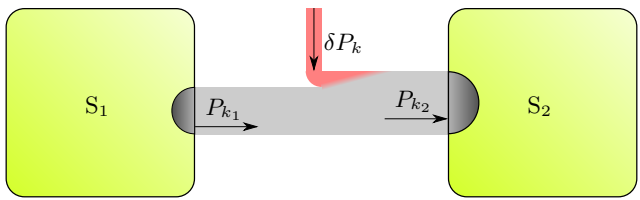
	\caption{%
		A residual power $\delta P_k = - (P_{k_1}+P_{k_2})$ emerges and distorts the dynamics of the full system when energy is exchanged between two simulators, S$_1$ and S$_2$, in a co\hyp{}simulation
	}
	\label{fig:power_bond_residual}
\end{figure}

The violation of energy conservation which we just uncovered in the coupling between two subsystems is, of course, an inherent issue with co\hyp{}simulation.
It stems from the fact that the simulators evolve their states independently of each other between communication time instances.
Luckily, it plays in our favor because it is also a measure of how accurate the co\hyp{}simulation solutions are at any given time.
In order to demonstrate this, let us first define a \emph{residual power}\footnote{%
Note that the sign convention for the residual power is different than the one used for energy fluxes encountered so far.
A positive residual power means that the energy of the total system increases.
}
\begin{equation}
\label{equ:residualpower}
	-
	(
	P_{k_1}
	+
	P_{k_2}
	)
	\equiv
	\delta P_k
\end{equation}
for the power bond $k$, see Fig.~\ref{fig:power_bond_residual} for an illustration.
If we construct input and output vectors such that
\begin{subequations}
\label{equ:residual_vectors}
\begin{equation}
\label{equ:coupling_vectors}
	\tilde{\vect{u}}_k
	\equiv
	\begin{pmatrix}
		\mbox{$\tilde{u}$}_{k_1}\\
		\mbox{$\tilde{u}$}_{k_2}
	\end{pmatrix}
	,
	\quad
	\vect{y}_k
	\equiv
	\begin{pmatrix}
		y_{k_1}\\
		y_{k_2}
	\end{pmatrix}
	,
\end{equation}
the residual power can be conveniently calculated from their scalar product,
\begin{equation}
\label{equ:residualpower_vectors}
	\delta P_k
	=
	-
	\tilde{\vect{u}}_k
	\cdot
	\vect{y}_k
	.
\end{equation}
\end{subequations}
If the bond between the simulators obeys energy conservation between communication points, the input and the output vector are orthogonal to each other and the residual power vanishes, $\delta P_k = 0$.
As already mentioned, this will not be the case in general, and energy either leaks from ($\delta P_k < 0$) or accumulates at ($\delta P_k > 0$) the power bond.
Let us now calculate a corresponding \emph{residual energy}
\begin{equation}
\label{equ:residualenergy}
	\delta E_k(t_{i+1})
	\equiv
	\int_{t_i}^{t_{i+1}}
	\delta P_k(t)
	\diff t
	,
\end{equation}
which tells us how much energy was incorrectly created due to the independent time integrations of the simulators during the macro time step $t_i \rightarrow t_{i+1} = t_i + \Delta t_i$.
When introduced into the total coupled system, these residual energy fluxes inevitably lead to altered energy densities and dissipation inside the subsystems due to the fact that energy is locally continuous, as expressed by Eq.~\eqref{eq:continuity_equation_power}.
They will, in other words, distort the dynamics of the system under consideration and in general decrease the accuracy of the co\hyp{}simulation and challenge its stability.
This is a profound statement as it tells us how much energy is \emph{wrongfully} added to the total energy of the overall coupled system and provides an intuitive explanation of why a co\hyp{}simulation produces inaccurate results or diverging solutions.

Let us for now assume constant extrapolation of the input values.
The residual power for the power bond $k$ at the communication point $t_{i+1}$ then becomes
\begin{subequations}
\label{equ:residuals_constant}
\begin{equation}
\label{equ:residualpower_constant}
	\delta P_k(t_{i+1})
	=
	-
	\vect{u}_k(t_i)
	\cdot
	\vect{y}_k(t_{i+1})
	.
\end{equation}
When calculating the residual energy, it is usually sufficient to make use of the rectangle quadrature rule,
\begin{equation}
\label{equ:residualenergy_constant}
	\delta E_k(t_{i+1})
	\approx
	\delta P_k(t_{i+1})
	\Delta t_i
	.
\end{equation}
\end{subequations}
In principle, higher-order corrections to the residual powers and energies can easily be calculated if the simulators also output (and input) higher-order derivatives along with the base values.
This, however, depends on the individual simulator implementation and it may be very impractical or even impossible to retrieve Jacobians.
If, on the other hand, inputs are approximated inside a simulator in terms of previous values and without the use or knowledge of derivatives (e.g.\ by use of Lagrange or Hermite polynomials \cite{Busch2012}, or Newton series), this will generally be unknown at the co\hyp{}simulation level.
Consequently, Eqs.~\eqref{equ:residuals_constant} are a fair choice to calculate residual energies on a general basis, and the forms provided here should therefore constitute a sufficient and practical approximation.
We will continue under the assumption of constant input extrapolation (i.e., the input values are held constant during the time integrations) for this reason.

Note that we can generalize Eqs.~\eqref{equ:residual_vectors} and group inputs and outputs together according to power bonds,
\begin{subequations}
\label{equ:residual_generalized}
\begin{equation}
\label{equ:residual_vectors_generalized}
	\tilde{\vect{u}}
	=
	\begin{pmatrix}
		\mbox{$\tilde{u}$}_{a_1}\\
		\mbox{$\tilde{u}$}_{a_2}\\
		\mbox{$\tilde{u}$}_{b_1}\\
		\mbox{$\tilde{u}$}_{b_2}\\
		\vdots
	\end{pmatrix}
	,
	\quad
	\vect{y}
	=
	\begin{pmatrix}
		y_{a_1}\\
		y_{a_2}\\
		y_{b_1}\\
		y_{b_2}\\
		\vdots
	\end{pmatrix}
	,
\end{equation}
such that the total residual power of all power bonds $\{a, b, \dots\}$ of the co\hyp{}simulation is obtained from
\begin{equation}
\label{equ:residualpower_vectors_generalized}
	\delta P
	=
	\sum_{k \in \{a, b, \dots\}}
	\delta P_k
	=
	-
	\tilde{\vect{u}}
	\cdot
	\vect{y}
	.
\end{equation}
\end{subequations}
Consequently, energy conservation is satisfied throughout the entire co\hyp{}simulation if the input and output vectors~\eqref{equ:residual_vectors_generalized} are orthogonal to each other.
On the other hand, Eqs.~\eqref{equ:residual_generalized} allows to easily calculate the total residual energy which is incorrectly added to the full coupled system due to co\hyp{}simulation coupling errors.

The concept of inaccurate energy transactions discussed here is closely related to the theoretical footing of the Nearly Energy Preserving Coupling Element (NEPCE) described in Ref.~\onlinecite{Benedikt2013}.
There, the local error of a simulator input and the resulting local error in the transmitted (generalized) power are discussed, albeit with sequential time integrations in the coupled simulators in mind.
In the present section, we aimed to cast a different light on the issue focusing in on parallel executions of coupled simulators and energy conservation throughout the entire coupled system.

We also intend to promote the use of power bonds for simulator coupling:
Power and energy transactions and their accuracies then become directly accessible with nothing more than the knowledge of the coupling variable values.
In this work, we only focus on the energetic coupling between two simulators.
The entire formalism along with all statements and results can easily be extended to direct couplings between a general number of simulators.
This is most effectively done by use of bond-graph-like junctions, and also keeps the modular character of co\hyp{}simulations intact.

\section{Error Estimation and Adaptive Step Size Control}
\label{sec:error_and_step_size}

In Section~\ref{sec:powerbonds}, we have seen how the fact that subsystems evolve their states independently of each other between discrete communication instances leads to the violation of energy conservation on the co\hyp{}simulation level.
The flow of energy between coupled simulators was discussed and the idea of power and energy residuals as a manifestation of this violation was introduced.
It is now time to reap the fruits of our efforts and apply these concepts to introduce a novel co\hyp{}simulation global error estimator without the need to repeat macro time steps and without any knowledge about the simulator implementations.
With it, we shall further propose a non-iterative adaptive macro step size controller based on energy conservation.
But first, let us review how global errors are usually estimated for co\hyp{}simulations.

\subsection{Local Error Estimators}
\label{subsec:error_estimator}

For simulator coupling without algebraic loops the glo\-bal error is bounded in terms of local errors.\cite{Arnold2013}
This allows us to find global error estimates in terms of local errors as a direct extension of ODE and DAE integration methods.
Error estimates are commonly based on comparing two approximations, $\tilde{\vect{y}}(t_{i+1})$ and $\tilde{\tilde{\vect{y}}}(t_{i+1})$, for the initial value problem, $\vect{y}' = \vect{f}(\vect{y},t)$, $\vect{y}(t_0) = \vect{y}_0$, at time $t_{i+1}$ in terms of an already obtained approximate solution at times $t \in \{t_0, t_1, \dots, t_i\}$.
The difference between the two approximations $\tilde{\vect{y}} - \tilde{\tilde{\vect{y}}}$ can then be used to estimate local errors and, consequently, derive a global error estimator.
Once a suitable estimator is found, it can also be used for adaptive control of the step size to obtain a balance between accuracy on one hand and efficiency on the other.

In a co\hyp{}simulation, there is naturally no access to the subsystem equations and the state variables are unavailable for local error estimation.
Instead, co\hyp{}simulation error estimators typically try to give an estimate of all numerical errors in the simulator \emph{outputs} when solving Eqs.~\eqref{equ:cosimulation} with approximated inputs.
Commonly, classic error estimation approaches are adapted directly for co\hyp{}simulation,\footnote{%
For a more comprehensive overview see, for example, Ref.~\citenum{Busch2012} and references therein.
}
and two approximations for the simulator outputs, $\tilde{\vect{y}}$ and $\tilde{\tilde{\vect{y}}}$, have to be obtained by carrying out a macro step twice:
An adaptive step size control by Richardson extrapolation \cite{Schierz2012,Arnold2013} is realized by carrying out a larger step $t_i \rightarrow t_{i+2}$ with step size $2 \Delta t$ and two smaller steps $t_i \rightarrow t_{i+1} \rightarrow t_{i+2}$ with step size $\Delta t$ and comparing the resulting output vectors.
In an embedded methods approach \cite{Guenther2001}, the integrations are performed twice with input polynomials of varying degrees.
Another approach \cite{Verhoeven2008} involves performing a compound macro step for two simulators and repeating the same step for the stiffer of the two subsystems using the updated coupling variables obtained from the first step.
Again, the difference between both results yields an error estimate.
Finally, error estimation based on Milne's device \cite{Busch2012} compares a predicted with a corrected solution, both of the same polynomial order.
In practice, all of these approaches suffer from one or several of the following shortcomings, and are not easily implemented for coupling with commercial simulation tools:
\begin{enumerate}
	\item They all require rollback, that is the ability to redo a macro step.
	This is often either not supported at all or complicated to realize in practice.
	Rollback is especially prohibited for real-time applications, such as Hardware-in-the-Loop.
	\item In addition, re-stepping is a time-consuming procedure, especially when performed at every communication point.
	Consequently, it somewhat mitigates the efficiency gain from adaptive step size control.
	\item Lastly, information essential to the error estimator method (for example, the internal time integration order) is often not available.
	Moreover, this requirement may expose sensitive information whose concealment is one of co\hyp{}simulation's strong suits, at least from an industrial perspective.
\end{enumerate}
The error estimator introduced by Busch \emph{et~al}.~\cite{Busch2011,Busch2012} is an exception worth mentioning because it requires neither re\hyp{}stepping nor internal simulator information.
We will discuss it further in Sections~\ref{subsec:error_estimator_Busch} and \ref{sec:comparison_busch}.

\subsection{Error Estimation Based on Energy Conservation}
\label{subsec:error_estimator_power}

As we have seen, co\hyp{}simulation coupling errors are directly expressed in terms of residual energies when pow\-er bonds are used.
These energy residuals give precisely the amount of energy that is incorrectly added to the total energy of the overall coupled system, and should naturally be contained in order to ensure co\hyp{}simulation accuracy and stability.
Because of this, they are well suited as versatile energy\hyp{}based error estimators.

Consider then the local error with respect to the power output of power port $k_\alpha$ for the macro time step $t_i \rightarrow t_{i+1} = t_i + \Delta t_i$,
\begin{equation}
\label{equ:error_local_power}
	\Delta {P_k}_\alpha(t_{i+1})
	=
	P_{k_\alpha}(t_{i+1})
	-
	P^0_{k_\alpha}(t_{i+1})
	,
\end{equation}
where $P^0_{k_\alpha}$ and ${P_k}_\alpha$ are the exact solution and the co\hyp{}simulation result, respectively, where $P_{k_\alpha}(t_i) = P^0_{k_\alpha}(t_i)$ is assumed.
For input extrapolation of order $m$, the local error in the inputs is \cite{Arnold2007,Arnold2014} $\Delta {\vect{u}_k} = \mathcal{O}({\Delta t}^{m+1})$ for sufficiently smooth problems and constant step sizes $\Delta t_i = \Delta t$.
Then, from considering Eq.~\eqref{equ:cosimulation}, the local error in the states is $\Delta {\vect{x}_k} = \mathcal{O}({\Delta t}^{m+2})$ and, consequently, the local error in the outputs is $\Delta {\vect{y}_k} = \mathcal{O}({\Delta t}^{m+2})$ if there is no direct feed-through (i.e., no direct dependence on any inputs), and $\Delta {\vect{y}_k} = \mathcal{O}({\Delta t}^{m+1})$ otherwise.
Using Eqs.~\eqref{equ:energyflow} and \eqref{equ:error_local_power} then yields $\Delta {P_k}_\alpha = \mathcal{O}({\Delta t}^{m+1})$, irrespective of the presence of direct feed-through.

Let us now take the sum of the local errors for both power ports of a power bond $k$.
We obtain $\delta P_k = - (\Delta P_{k_1} + \Delta P_{k_2})$, where we used Eqs.~\eqref{eq:continuity_equation_powerbond} and \eqref{equ:residualpower}.
Therefore, the residual power gives the average of the local errors in the power,
\begin{subequations}
\label{equ:error_residual_power}
\begin{equation}
\label{equ:error_vs_residual_power}
	\overline{\Delta {P_k}}
	=
	-
	\frac{1}{2}
	\delta P_k
	.
\end{equation}
Similarly, the residual energy gives the average local error in the energy transmitted during the macro time step $t_i \rightarrow t_{i+1}$,
\begin{equation}
\label{equ:error_vs_residual_energy}
	\overline{\Delta {E_k}}(t_{i+1})
	=
	-
	\frac{1}{2}
	\delta E_k(t_{i+1})
	,
\end{equation}
\end{subequations}
where we carried out the time integration according to Eq.~\eqref{equ:residualenergy}.
Note that Eqs.~\eqref{equ:error_residual_power} are \emph{exact}, irrespective of macro or micro step sizes, and irrespective of the simulator implementations.

\begin{figure}[h!tb]
	\centering
	\includegraphics[width=\graphicswidth]{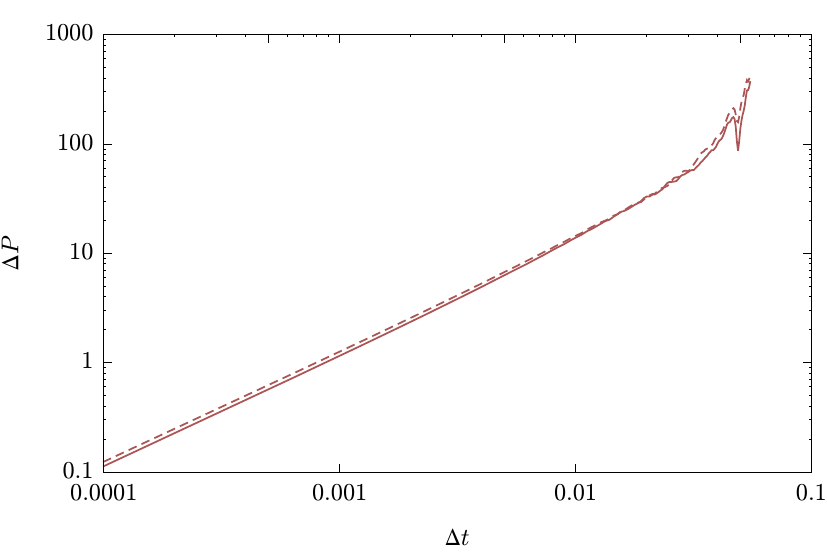}
	\caption{%
		Absolute error in the power (dashed) and power residual estimate (solid) as a function of the macro step size for the power bond in the linear benchmark model from Sec.~\ref{sec:benchmark_model}
	}
	\label{fig:error_estimator}
\end{figure}

The suitability of energy residuals as energy\hyp{}conservation\hyp{}based error estimators is exemplified in Fig.~\ref{fig:error_estimator} for the power bond in the co\hyp{}simulation benchmark model studied in Sec.~\ref{sec:benchmark_model}.
Shown is Eq.~\eqref{equ:error_vs_residual_power} averaged over the entire simulation time $T$,
\begin{equation*}
	\frac{1}{2}
	\overline{|\delta P_k|}
	=
	\frac{1}{2 T}
	\sum_i
	|\delta E_k(t_{i+1})|,
\end{equation*}
against the average error in the power
\begin{equation*}
	\overline{|\Delta P_k|}
	=
	\frac{1}{T}
	\sum_i
	|\Delta P_k(t_{i+1})|
	\Delta t_i,
\end{equation*}
where $\Delta P_k(t_{i+1}) = P_{k_{12}}(t_{i+1}) - P^0_{k_{12}}(t_{i+1})$, and $P^0_{k_{12}}$ is the exact solution.

Using energies as error metrics as opposed to non-energetic quantities (e.g.\ positions) has the advantage that energy considerations are directly taken into account as well and offers a more holistic approach.\cite{Gonzalez2011}
Outside the realm of co\hyp{}simulations, energy errors are used as a measure of quality in hybrid earthquake simulations \cite{Thewalt1994,Mosqueda2007} or in molecular dynamics simulations, for example.
Additionally, the method proposed here naturally solves the issue of the numerical values of the outputs lying on very different scales for force-displacement \cite{Busch2012,Schierz2012,Arnold2014} coupling, or the force-velocity coupling we shall investigate in the next section:
The outputs representing forces will typically have much larger numeric values than the ones representing displacements or velocities.
As a result, local errors from some simulators may be given weightings which are much too large compared to others.

\subsection{Adaptive Step Size Controller}
\label{subsec:step_controller}

Let us now define a scalar error indicator,
\begin{equation}
\label{equ:error_indicator}
	\epsilon(t)
	\equiv
	\sqrt{%
		\frac{1}{N}
		\sum_{k = 1}^N
		\bigg(%
			\frac{%
				\delta E_k(t)
			}{%
				r_k
				\big(
				{E_0}_k
				+
				| E_k(t) |
				\big)%
		}%
		\bigg)^2
	}
	,
\end{equation}
which contains the residual energies $\delta E_k$ and the energies $E_k(t_{i+1}) \approx P_{k_{12}}(t_{i+1}) \Delta t_i$ transmitted per time step for all $N$ power bonds.
Here, ${E_0}_k$ is a typical energy scale and $r_k$ a relative tolerance, respectively.
Both can be chosen individually per power bond and effectively determine the bond's energy resolution.
If $\epsilon \leq 1$, the error is sufficiently small with respect to the defined tolerances and energy scales.
Values of $\epsilon > 1$ indicate that the tolerances are exceeded and the co\hyp{}simulation result is potentially unreliable.
Applying, for example, a PI-controller \cite{Gustafsson1988,Burrage2004} determines a new optimal step size based on Eq.~\eqref{equ:error_indicator},
\begin{equation}
\label{equ:step_control}
	{\Delta t}_{i+1}
	=
	\alpha_\text{s}
	\epsilon(t_i)^{-k_\text{I}-k_\text{P}}
	\epsilon(t_{i-1})^{k_\text{P}}
	\Delta t_i
	,
\end{equation}
where $k_\text{I}$ and $k_\text{P}$ are the integral and the proportional gain, respectively, and $\alpha_\text{s} \in [0.8,0.9]$ is a safety factor.
We choose $k_\text{I} = 0.3/(m+2)$ and $k_\text{P} = 0.4/(m+2)$, where $m$ is the extrapolation order ($m=0$ for constant extrapolation).
It is beneficial to introduce user-defined bounds for the step size, ${\Delta t}_\text{min} \leq \Delta t_{i+1} \leq {\Delta t}_\text{max}$, as well as limit its rate of change, $\Theta_\text{min} \leq \Delta t_{i+1}/\Delta t_i \leq \Theta_\text{max}$. Typical parameter choices are $\Theta_\text{min} \in [0.2,0.5]$ and $\Theta_\text{max} \in [1.5,5.0]$.

\begin{table}[h!tb]
	\caption{%
		Configuration of the adaptive step size controller for the benchmark model in Sec.~\ref{sec:benchmark_model}
	}
	\label{tab:controller_configuration}
	\centering
	\begin{tabular}{lS[table-format=3.1]s}
		\hline\noalign{\smallskip}
		 & {Value} & {Unit} \\
		\noalign{\smallskip}\hline\noalign{\smallskip}
		$\alpha_\text{s}$ & 0.8 & \\
		${\Delta t}_\text{min}$ & 0.1 & \milli\second \\
		${\Delta t}_\text{max}$ & 10.0 & \milli\second \\
		$\Theta_\text{min}$ & 0.2 & \\
		$\Theta_\text{max}$ & 1.5 & \\
		$E_0$ & 750.0 & \si{\joule} \\
		\noalign{\smallskip}\hline
	\end{tabular}
\end{table}

We have thus defined a novel co\hyp{}simulation method for error estimation and adaptive step size control based on energy conservation considerations.
It does not require the repetition of macro time steps, nor does it demand any knowledge about simulator implementations.
In Section~\ref{sec:benchmark_model}, we will demonstrate how it can be used to make co\hyp{}simulations more accurate and more efficient.

\subsection{Predictor/Corrector Error Estimator}
\label{subsec:error_estimator_Busch}

As mentioned previously, another error estimation and step size control co\hyp{}simulation algorithm which is non\hyp{}iterative in nature is the one proposed by Busch \emph{et~al}.~\cite{Busch2011,Busch2012}.
It makes use of polynomial extrapolation of order $r = m + 1$ to construct a prediction $\tilde{\vect{y}}_k^{(r)}(t_{i+1})$ of the output vector~\eqref{equ:cosimulation_outputs} using the previously obtained values $\{ \vect{y}_k(t_i), \dots, \vect{y}_k(t_{i-r}) \}$.
The difference between this predictor and the actually obtained output vector---the corrector---gives a local error estimate,~\cite{Busch2011,Busch2012}
\begin{equation}
	\delta {\tilde{\vect{y}}_k}(t_{i+1})
	\equiv
	\vect{y}_k(t_{i+1})
	-
	\tilde{\vect{y}}_k^{(r)}(t_{i+1})
	\approx
	\Delta {\vect{y}_k}(t_{i+1})
	,
\end{equation}
and can be used for adaptive control of the macro step size:
A PI\hyp{}controller~\eqref{equ:step_control} with $k_\text{I} = 0.3/(m+1)$ and $k_\text{P} = 0.4/(m+1)$ proposes a new step size based on the scalar error indicator~\cite{Busch2011,Busch2012}
\begin{equation}
\label{equ:error_indicator_Busch}
	\tilde{\epsilon}(t)
	\equiv
	\max_\alpha
	\frac{1}{\text{TOL}}
	\frac{%
		y_{k_\alpha}(t)
		-
		\tilde{y}_{k_\alpha}^{(r)}(t)
	}{%
		1
		+
		\rho
		\max
		\big(
			|y_{k_\alpha}(t)|,
			|\tilde{y}_{k_\alpha}(t)|
		\big)
	}
	,
\end{equation}
where $\rho$ sets the weight between the relative and absolute errors and $\text{TOL}$ is a user-defined tolerance.

This predictor/corrector approach and the energy\hyp{}conservation\hyp{}based method (ECCO) discussed in the previous sections are both easily implemented for non-iterative co\hyp{}simulation error estimation and adaptive step size control.
But they are several noteworthy differences to be found between them:
\begin{itemize}
\item
	ECCO gives the \emph{exact} value of the average local error in the power outputs of two connected power ports, see Eqs.~\eqref{equ:error_residual_power}.
	This is true for any macro or micro step size, and irrespective of the simulator implementations.
	By contrast, the predictor/corrector method may be able to provide good estimates only for relatively small macro time step sizes and is generally sensitive to the polynomial order of the predictor.\cite{Busch2011,Busch2012}
\item
	ECCO is based on the use of power bonds, and while we firmly believe that these are generally a great choice for co-simulation---especially from an engineering and industrial perspective---it is also safe to assume that the vast majority of available models and tools does not use power bonds as of yet.
	In that sense, the predictor/corrector method is less demanding and likely more applicable given the status quo.
\item
	As mentioned in Section~\ref{subsec:error_estimator_power}, the use of powers and energies also means that ECCO's global error estimator~\eqref{equ:error_indicator} is insensitive to the scaling of simulator output values, while Eq.~\eqref{equ:error_indicator_Busch} is not.
	This can lead to practical difficulties in the implementation of the predictor/corrector method, as we shall see in Section~\ref{sec:comparison_busch}.
\end{itemize}

\section{Co\hyp{}Simulation Benchmark Model}
\label{sec:benchmark_model}

\begin{figure}[h!tb]
	\centering
	\def\svgwidth{\graphicswidth}
	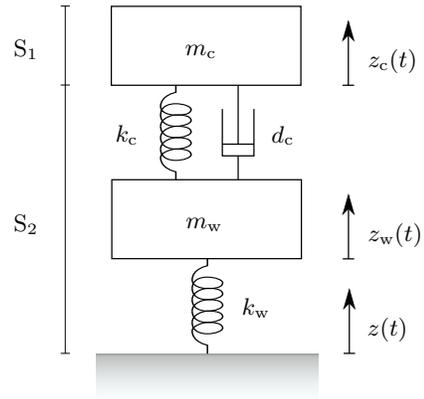
	\caption{%
		The quarter car benchmark model is split into the subsystems S$_1$ and S$_2$ for co\hyp{}simulation
	}
	\label{fig:quartercar_model}
\end{figure}

A generic test case is needed in order to examine and compare the performance of various co\hyp{}simulation methods.
The mechanical interpretation of Dahlquist's test equation for stability analysis of integration methods \cite{Dahlquist1956} is a linear 1-DOF oscillator.
Because it can be considered two coupled Dahlquist equations, the use of a linear 2-DOF oscillator as a co\hyp{}simulation test case has been proposed.\cite{Busch2011,Busch2012,Schierz2012,Arnold2013,Arnold2014,Gonzalez2011,Schweizer2014a,Schweizer2014b,Clauss2012}
This test system is also a simplification of numerous relevant real engineering systems.
One such example is the quarter car model \cite{Popp2010} depicted in Fig.~\ref{fig:quartercar_model}.
We too shall adopt it as a co\hyp{}simulation benchmark model and slightly modify it to include nonlinearity.

\begin{table}[h!tb]
	\caption{%
		Parameters for the linear quarter car benchmark model according to Ref.~\onlinecite{Arnold2013}
	}
	\label{tab:quartercar_model}
	\centering
	\begin{tabular}{lS[table-format=6.1]s}
		\hline\noalign{\smallskip}
		 & {Value} & Unit \\
		\noalign{\smallskip}\hline\noalign{\smallskip}
		$m_\text{c}$ & 400.0 & \kilogram \\
		$m_\text{w}$ & 40.0 & \kilogram \\
		$k_\text{c}$ & 15000.0 & \newton\per\meter \\
		$k_\text{w}$ & 150000.0 & \newton\per\meter \\
		$d_\text{c}$ & 1000.0 & \newton\second\per\meter \\
		$n_d$ & 0.5 & \\
		\noalign{\smallskip}\hline
	\end{tabular}
\end{table}

\begin{figure}[h!tb]
	\centering
	\includegraphics[width=\graphicswidth]{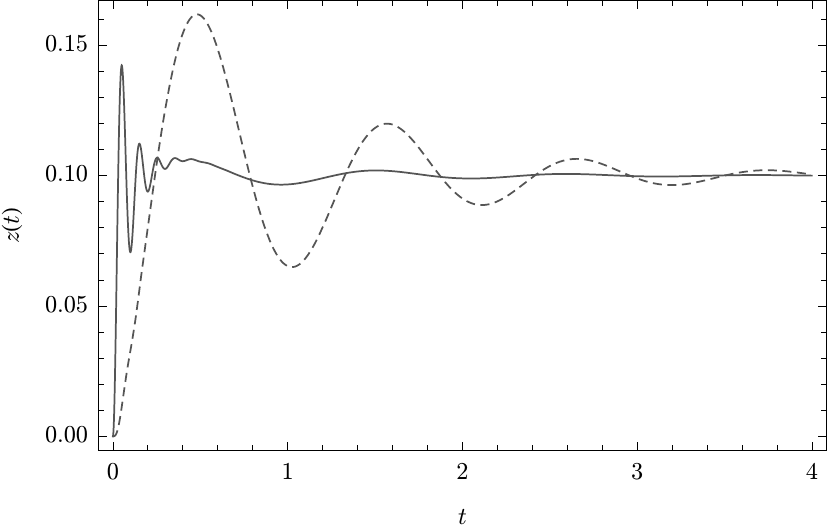}
	\caption{%
		Displacements of wheel (solid) and chassis (dashed) for the linear quarter car benchmark
	}
	\label{fig:quartercar_positions}
\end{figure}

The equations of motion describe the displacements $z_\text{c}$ and $z_\text{w}$ of the masses $m_\text{c}$ and $m_\text{w}$ of the chassis and the wheel, respectively, according to
\begin{subequations}
\label{equ:quartercar}
\begin{gather}
\label{equ:quartercar_S1}
	m_\text{c} \ddot{z}_\text{c}(t)
	=
	-
	F_\text{c}(t)
	,
	\\
\label{equ:quartercar_S2}
	m_\text{w} \ddot{z}_\text{w}(t)
	=
	-
	F_\text{w}(t)
	+
	F_\text{c}(t)
	.
\end{gather}
The spring-damper forces are given by \cite{Busshardt1992,Arnold2013}
\begin{gather}
\label{equ:quartercar_Fc}
\begin{split}
	F_\text{c}(t)
	&=
	k_\text{c}
	\big(%
		z_\text{c}(t)
		-
		z_\text{w}(t)
	\big)
	\\
	&+
	d_\text{c}
	\sign
	\big(%
		\dot{z}_\text{c}(t)
		-
		\dot{z}_\text{w}(t)
	\big)
	\big|%
		\dot{z}_\text{c}(t)
		-
		\dot{z}_\text{w}(t)
	\big|
	^
	{2/(1+2 n_d)}
	,
\end{split}	
	\\
\label{equ:quartercar_Fw}
	F_\text{w}(t)
	=
	k_\text{w}
	(%
		z_\text{w}(t)
		-
		z(t)
	)
	,
\end{gather}
\end{subequations}
where $k_\text{c}$ and $k_\text{w}$ are the spring constants, $d_\text{c}$ is the damping constant, and $n_d$ tunes the linearity of the damping force.
The parameter values are chosen as list\-ed in Table~\ref{tab:quartercar_model}.
The system is excited with the function
\begin{align}
	z(t)
	=
	\begin{cases}
		0, \quad &t < 0,\\
		0.1, \quad &t \geq 0,
	\end{cases}
\end{align}
and the reference solution for the displacements of chassis and wheel with the initial conditions $z_\text{c} = z_\text{w} = \dot{z}_\text{c} = \dot{z}_\text{w} = 0$ at $t=0$ is shown in Fig.~\ref{fig:quartercar_positions}.

For the co\hyp{}simulation, the system is split into two simulators, see Fig.~\ref{fig:quartercar_model}:
\begin{enumerate}
	\item[S$_1$] models the chassis only, described by Eq.~\eqref{equ:quartercar_S1} with the coupling variables
		\begin{subequations}
		\label{equ:quartercar_coupling}
		\begin{align}
		\label{equ:quartercar_coupling_S1}
			u_1(t_i)
			&=
			-
			F_\text{c}(t_i)
			,&
			y_1(t_{i+1})
			&=
			\dot{z}_\text{c}(t_{i+1})
			.
\intertext{\item[S$_2$] contains the remainder of the system and is describ\-ed by Eq.~\eqref{equ:quartercar_S2} with the coupling variables}
		\label{equ:quartercar_coupling_S2}
			u_2(t_i)
			&=
			\dot{z}_\text{c}(t_i)
			,&
			y_2(t_{i+1})
			&=
			F_\text{c}(t_{i+1})
			.
		\end{align}
		\end{subequations}
\end{enumerate}
The connections~\eqref{equ:quartercar_coupling} define a power bond and we can apply the residual energy concept.
We use the forward Euler method to carry out the time integration in S$_2$ with micro step sizes of ${\Delta t}_{\text{S}_2} = {\Delta t}/10$, where ${\Delta t}$ is the macro step size.\footnote{%
S$_1$ can be solved exactly with constant extrapolation of the input $F_\text{c}(t) = F_\text{c}(t_i)$ for $t \in (t_i,t_{i+1}]$.
}

\begin{figure*}[h!tb]
	\centering
	\includegraphics[width=\graphicswidthfull]{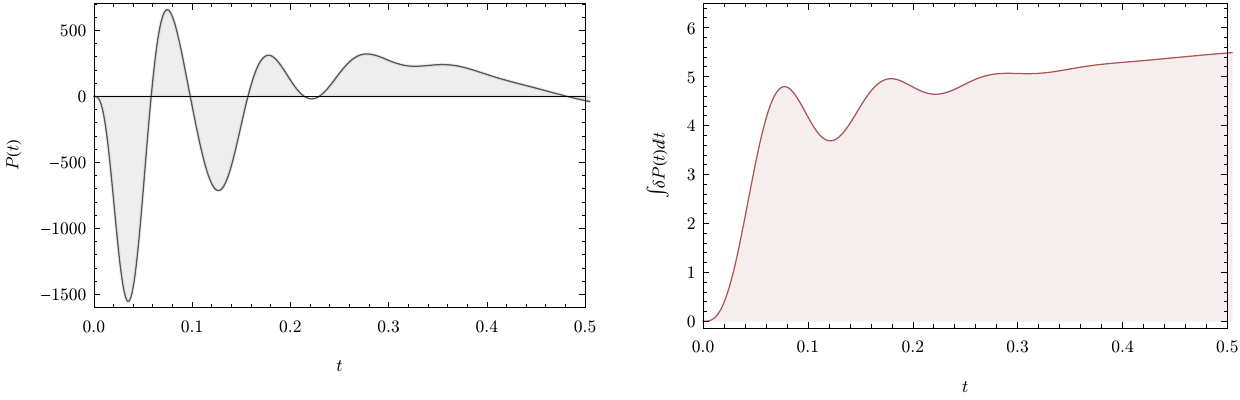}
	\caption{%
		Energy transactions for the linear quarter car benchmark with a constant macro step size $\Delta t = \SI{1}{\milli\second}$:
		power transmitted from S$_1$ to S$_2$ (left) and the residual energy accumulated over time (right)
	}
	\label{fig:quartercar_energy}
\end{figure*}

The energy transactions between the simulators are plotted in Fig.~\ref{fig:quartercar_energy} for a constant macro step size of $\Delta t = \SI{1}{\milli\second}$:
Energy is first transmitted from S$_2$ to S$_1$ as the potential energy stored in $k_\text{w}$ is transformed into kinetic energy of $m_\text{c}$.
However, these energy transfers are not accurate, and residual energy is accumulated over the power bond and added to the total energy of the overall system.
This way, a total of $\int \delta P(t) \diff t \approx \SI{6.4}{\joule}$ is incorrectly added to the coupled system during the entire simulation.
By comparison, the total amount of energy initially stored in the spring $k_\text{w}$ and finally entirely dissipated in the damper $d_\text{c}$ is $E_0 = \SI{750}{\joule}$.

\subsection{Adaptive Step Size Control}
\label{subsec:adaptive_step_size}

Let us now demonstrate the performance of the adaptive step size controller as described in Section~\ref{sec:error_and_step_size} to improve co\hyp{}simulation accuracy and efficiency.
The residual energy concept is used to propose an optimal macro step size for the \emph{next} time step in order to minimize residual energies and hence increase accuracy.
As an example, consider the previous case of the linear quarter car model.
We use the scalar error indicator \eqref{equ:error_indicator} and configure the PI-controller \eqref{equ:step_control} according to the parameters listed in Table~\ref{tab:controller_configuration}, with a starting step size ${\Delta t}_0 = {\Delta t}_\text{min}$.
The energy scale is set to the initial energy excitation $E_0 = \SI{750}{\joule}$ of the system, and the balance between accuracy and computational efficiency can conveniently be tuned with the relative tolerance~$r$.

\begin{table}[h!tb]
	\caption{%
		Linear quarter car benchmark results with constant step size and residual\hyp{}energy\hyp{}based adaptive step size control
	}
	\label{tab:quartercar_results}
	\centering
	\begin{tabular}{lS[table-format=1.1d-1]S[table-format=1.1]S[table-format=1.1]S[table-format=1.1]S[table-format=1.1]}
		\hline\noalign{\smallskip}
  		\multicolumn{1}{c}{} & \multicolumn{2}{c}{Algorithm} & \multicolumn{1}{ c }{Power} & \multicolumn{2}{ c }{Error} \\
 		\multicolumn{1}{ c }{} & \multicolumn{1}{ c }{tolerance} & \multicolumn{1}{ c }{$\frac{\overline{\Delta t}}{\si{\milli\second}}$} & \multicolumn{1}{ c }{$\frac{\overline{P_{12}}}{\si{\watt}}$} & \multicolumn{1}{ c }{$\frac{\overline{|\Delta P|}}{\si{\watt}}$} & \multicolumn{1}{ c}{$\frac{\smallint\!\delta P}{\si{\joule}}$} \\
		\noalign{\smallskip}\hline\noalign{\smallskip}
		constant & & 1. & 0.4 & 1.3 & 6.4 \\
		\noalign{\smallskip}\hline\noalign{\smallskip}
		adaptive & 2.8d-6 & 1. & 0.0 & 0.4 & 1.6 \\
		\noalign{\smallskip}\hline\noalign{\smallskip}
		adaptive & 3.1d-5 & 2.9 & 0.1 & 1.3 & 5.0 \\
		\noalign{\smallskip}\hline
	\end{tabular}
\end{table}

\begin{figure}[h!tb]
	\centering
	\includegraphics[width=\graphicswidth]{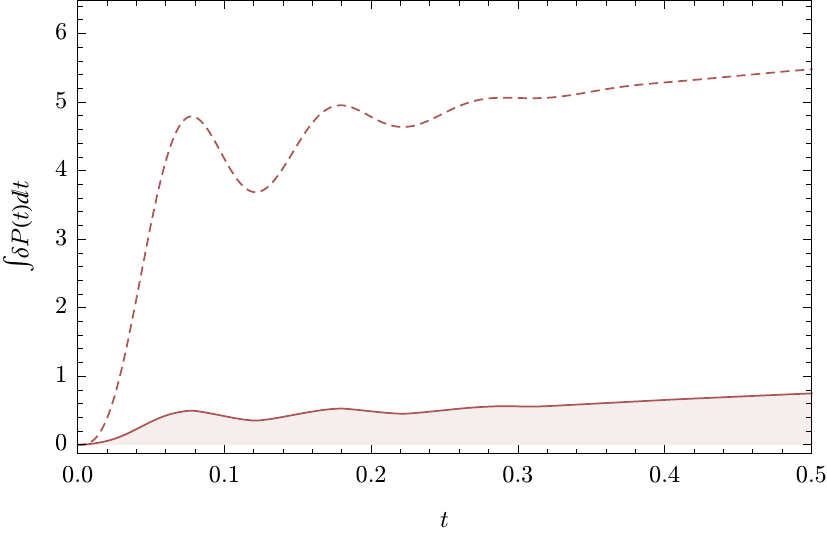}
	\caption{%
		Accumulation of residual energy for the linear quarter car benchmark:
		Adaptive step size control with $\overline{\Delta t} \approx \SI{1}{\milli\second}$ (solid) and constant step size $\Delta t = \SI{1}{\milli\second}$ (dashed)
	}
	\label{fig:quartercar_adaptive}
\end{figure}

Let us first demonstrate the improvements in accuracy and set the tolerance to $r = \num{2.8d-6}$ such that the average macro step size is $\overline{\Delta t} \approx \SI{1}{\milli\second}$.
While keeping the calculation time approximately the same, the mean absolute error in the power $\overline{|\Delta P|}$ is reduced by \SI{70}{\percent} relative to the result with constant step size, see Table~\ref{tab:quartercar_results}.
As Fig.~\ref{fig:quartercar_adaptive} exemplifies, the residual energy is kept significantly better contained with adaptive step size control:
A total of $\int \delta P(t) \diff t \approx \SI{1.6}{\joule}$ is added to the coupled system during the entire simulation, only \SI{25}{\percent} of the previous value.

Next, we increase the relative tolerance such that the mean absolute error in the power is approximately the same as with a constant step size of $\Delta t = \SI{1}{\milli\second}$.
With $r = \num{3.1d-5}$, a value of $\overline{|\Delta P|} \approx \SI{1.3}{\watt}$ is now reached with only about $1/3$ of the total time steps required without the algorithm.
Energy flows are still described more accurately than with constant step sizes, and the total residual energy accumulated during the simulation run is $\int \delta P(t) \diff t \approx \SI{5.0}{\joule}$.

\subsection{Nonlinear Damping}
\label{subsec:nonlinear_spring}

\begin{figure}[h!tb]
	\centering
	\includegraphics[width=\graphicswidth]{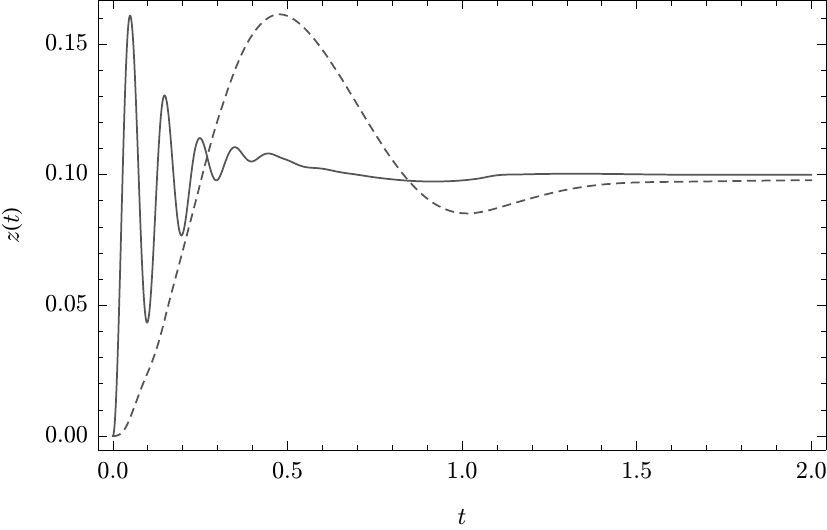}
	\caption{%
		Displacements of wheel (solid) and chassis (dashed) for the nonlinear quarter car benchmark
	}
	\label{fig:quartercar_nonlinear_positions}
\end{figure}

Let us now investigate the effects of nonlinear damping on the system dynamics and co\hyp{}simulation accuracy.
To this end, we modify the parameters in Eq.~\eqref{equ:quartercar_Fc} according to Table~\ref{tab:quartercar_nonlinear_model}.
The reference solution for the displacements of chassis and wheel is shown in Fig.~\ref{fig:quartercar_nonlinear_positions}.
Note that both masses return to their equilibrium positions much faster than in the previous case with linear damping.
Because of this, we only simulate the system up to $t = \SI{2}{\second}$.

\begin{table}[h!tb]
	\caption{%
		Parameter changes to include nonlinear damping forces in the benchmark model according to Ref.~\onlinecite{Busshardt1992}
	}
	\label{tab:quartercar_nonlinear_model}
	\centering
	\begin{tabular}{lS[table-format=3.1]s}
		\hline\noalign{\smallskip}
		 & {Value} & {Unit} \\
		\noalign{\smallskip}\hline\noalign{\smallskip}
		$d_\text{c}$ & 900.0 & \newton\raiseto{1/2}{(\second\per\meter)} \\
		$n_d$ & 1.5 & \\
		\noalign{\smallskip}\hline
	\end{tabular}
\end{table}

\begin{figure*}[h!tb]
	\centering
	\includegraphics[width=\graphicswidthfull]{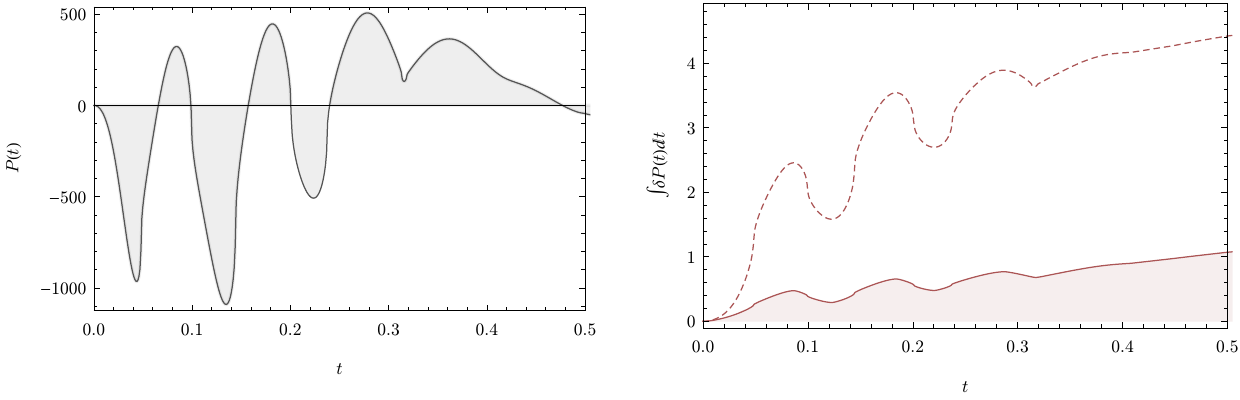}
	\caption{%
		Energy transactions for the nonlinear benchmark:
		Power transmitted from S$_1$ to S$_2$ with a constant macro step size $\Delta t = \SI{1}{\milli\second}$ (left), and the residual energy accumulated over time for adaptive control of the macro step size with $\overline{\Delta t} \approx \SI{1}{\milli\second}$ (right, solid) and with a constant step size $\Delta t = \SI{1}{\milli\second}$ (right, dashed)
	}
	\label{fig:quartercar_nonlinear_energy}
\end{figure*}

As was the case with linear damping, using the energy\hyp{}based adaptive step size control significantly improves accuracy.
Table~\ref{tab:quartercar_nonlinear_results} lists all results discussed here.
With the tolerance set to $r = \num{7.5d-6}$, the computational cost is kept about the same as with constant step sizes ($\overline{\Delta t} \approx \SI{1}{\milli\second}$) while the errors are minimized and energy flows described more accurately, as demonstrated by Fig.~\ref{fig:quartercar_nonlinear_energy}.
The mean absolute error in the power and the total residual energy wrongfully added to the coupled system during simulation time are both reduced by \SI{70}{\percent} with step size control.
Increasing the relative tolerance to $r = \num{1.0d-4}$ keeps the mean absolute error in the power approximately the same but improves computational efficiency:
It only takes about $1/3$ of the time steps to obtain a value of $\overline{|\Delta P|} \approx \SI{4}{\watt}$.

\begin{table}[h!tb]
	\caption{%
		Nonlinear benchmark results with constant step size and residual\hyp{}energy\hyp{}based adaptive step size control
	}
	\label{tab:quartercar_nonlinear_results}
	\centering
	\begin{tabular}{lS[table-format=1.1d-1]S[table-format=1.1]S[table-format=1.1]S[table-format=1.1]S[table-format=1.1]}
		\hline\noalign{\smallskip}
		\multicolumn{1}{c}{} & \multicolumn{2}{c}{Algorithm} & \multicolumn{1}{ c }{Power} & \multicolumn{2}{ c }{Error} \\
		\multicolumn{1}{ c }{} & \multicolumn{1}{ c }{tolerance} & \multicolumn{1}{ c }{$\frac{\overline{\Delta t}}{\si{\milli\second}}$} & \multicolumn{1}{ c }{$\frac{\overline{P_{12}}}{\si{\watt}}$} & \multicolumn{1}{ c }{$\frac{\overline{|\Delta P|}}{\si{\watt}}$} & \multicolumn{1}{ c}{$\frac{\smallint\!\delta P}{\si{\joule}}$} \\
		\noalign{\smallskip}\hline\noalign{\smallskip}
		constant & & 1. & 1 & 4 & 5 \\
		\noalign{\smallskip}\hline\noalign{\smallskip}
		adaptive & 7.5d-6 & 1. & 0.0 & 1.1 & 1.6 \\
		\noalign{\smallskip}\hline\noalign{\smallskip}
		adaptive & 1.0d-4 & 3.1 & 0 & 4 & 6 \\
		\noalign{\smallskip}\hline
	\end{tabular}
\end{table}

\subsection{Alternative System Reticulation}
\label{subsec:alt_reticulation}

\begin{figure}[h!tb]
	\centering
	\def\svgwidth{\graphicswidth}
	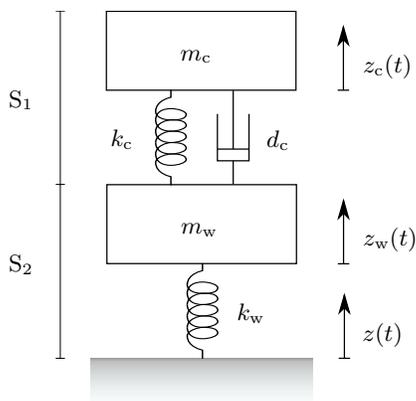
	\caption{%
		Alternative reticulation of the quarter car benchmark model with subsystems S$_1$ and S$_2$
	}
	\label{fig:quartercar_model_alt}
\end{figure}

The way a system is reticulated can have a substantial influence on accuracy, stability, and efficiency of a co\hyp{}simulation.
To exemplify this and further demonstrate the applicability and performance of the adaptive step size controller described in Sec.~\ref{sec:error_and_step_size}, we consider the quarter car model with a different subsystem splitting, see Fig.~\ref{fig:quartercar_model_alt}:
\begin{enumerate}
	\item[S$_1$] is described by Eq.~\eqref{equ:quartercar_S1} but now also includes the spring-damper element \eqref{equ:quartercar_Fc} with the coupling variables
		\begin{subequations}
		\label{equ:quartercar_alt_coupling}
		\begin{align}
		\label{equ:quartercar_alt_coupling_S1}
			u_1(t_i)
			&=
			\dot{z}_\text{w}(t_i)
			,&
			y_1(t_{i+1})
			&=
			F_\text{c}(t_{i+1})
			.
\intertext{\item[S$_2$] is described by Eq.~\eqref{equ:quartercar_S2} with the coupling variables}
		\label{equ:quartercar_alt_coupling_S2}
			u_2(t_i)
			&=
			-
			F_\text{c}(t_i)
			,&
			y_2(t_{i+1})
			&=
			\dot{z}_\text{w}(t_{i+1})
			.
		\end{align}
		\end{subequations}
\end{enumerate}
The connections~\eqref{equ:quartercar_alt_coupling} define a power bond and the residual energy concept can be applied.
This time, time integration needs to be carried out in both simulators, and we initially choose micro step sizes of ${\Delta t}_{S_1} ={\Delta t}_{S_2} = {\Delta t}/10$.
Again, the forward Euler method is used in both simulators.

\begin{table}[h!tb]
	\caption{%
		Linear benchmark results for alternative system reticulation with constant step size and residual\hyp{}energy\hyp{}based adaptive step size control
	}
	\label{tab:quartercar_alt_results}
	\centering
	\begin{tabular}{lS[table-format=1.1d-1]S[table-format=1.1]S[table-format=-1.3]S[table-format=1.3]S[table-format=1.3]}
		\hline\noalign{\smallskip}
		\multicolumn{1}{c}{} & \multicolumn{2}{c}{Algorithm} & \multicolumn{1}{ c }{Power} & \multicolumn{2}{ c }{Error} \\
		\multicolumn{1}{ c }{} & \multicolumn{1}{ c }{tolerance} & \multicolumn{1}{ c }{$\frac{\overline{\Delta t}}{\si{\milli\second}}$} & \multicolumn{1}{ c }{$\frac{\overline{P_{12}}}{\SI{d2}{\watt}}$} & \multicolumn{1}{ c }{$\frac{\overline{|\Delta P|}}{\SI{d2}{\watt}}$} & \multicolumn{1}{ c}{$\frac{\smallint\!\delta P}{\SI{d2}{\joule}}$} \\
		\noalign{\smallskip}\hline\noalign{\smallskip}
		constant & & 1. & -1.92 & 0.12 & 0.23 \\
		\noalign{\smallskip}\hline\noalign{\smallskip}
		adaptive & 9.1d-7 & 1. & -1.879 & 0.013 & 0.016 \\
		\noalign{\smallskip}\hline
	\end{tabular}
\end{table}

As can be seen from the simulation results listed in Tables~\ref{tab:quartercar_alt_results} and \ref{tab:quartercar_alt_nonlinear_results}, this system reticulation is much less favorable than the one discussed previously, giving larger errors for linear and nonlinear damping alike.
The co\hyp{}simulation is also less stable:
With linear damping, for example, the previous reticulation became unstable around a constant step size of $\Delta t \approx \SI{58.5}{\milli\second}$, while instability already sets in at around $\Delta t \approx \SI{11.3}{\milli\second}$ with the reticulation discussed here.
The effects of such an ill-chosen reticulation can be mitigated and the accuracy enhanced substantially by employing the adaptive step size controller.
The errors are then reduced by between \SI{80}{\percent} and \SI{93}{\percent} at no additional computational cost.

\begin{table}[h!tb]
	\caption{%
		Nonlinear benchmark results for alternative system reticulation with constant step size and residual\hyp{}energy\hyp{}based adaptive step size control
	}
	\label{tab:quartercar_alt_nonlinear_results}
	\centering
	\begin{tabular}{lS[table-format=1.1d-1]S[table-format=1.1]S[table-format=-1.2]S[table-format=1.2]S[table-format=1.2]}
		\hline\noalign{\smallskip}
		\multicolumn{1}{ c }{} & \multicolumn{2}{c}{Algorithm} & \multicolumn{1}{ c }{Power} & \multicolumn{2}{ c }{Error} \\
		\multicolumn{1}{ c }{} & \multicolumn{1}{ c }{tolerance} & \multicolumn{1}{ c }{$\frac{\overline{\Delta t}}{\si{\milli\second}}$} & \multicolumn{1}{ c }{$\frac{\overline{P_{12}}}{\SI{d2}{\watt}}$} & \multicolumn{1}{ c }{$\frac{\overline{|\Delta P|}}{\SI{d2}{\watt}}$} & \multicolumn{1}{ c}{$\frac{\smallint\!\delta P}{\SI{d2}{\joule}}$} \\
		\noalign{\smallskip}\hline\noalign{\smallskip}
		constant & & 1. & -3.9 & 0.3 & 0.5 \\
		\noalign{\smallskip}\hline\noalign{\smallskip}
		adaptive & 2.4d-5 & 1. & -3.77 & 0.05 & 0.05 \\
		\noalign{\smallskip}\hline
	\end{tabular}
\end{table}

\begin{table}[h!tb]
	\caption{%
		Linear benchmark results for alternative system reticulation with a low accuracy time integration in S$_2$
	}
	\label{tab:quartercar_alt_results_lowS2}
	\centering
	\begin{tabular}{lS[table-format=1.1d-1]S[table-format=1.1]S[table-format=-1.2]S[table-format=1.2]S[table-format=1.2]}
		\hline\noalign{\smallskip}
		\multicolumn{1}{ c }{} & \multicolumn{2}{c}{Algorithm} & \multicolumn{1}{ c }{Power} & \multicolumn{2}{ c }{Error} \\
		\multicolumn{1}{ c }{} & \multicolumn{1}{ c }{tolerance} & \multicolumn{1}{ c }{$\frac{\overline{\Delta t}}{\si{\milli\second}}$} & \multicolumn{1}{ c }{$\frac{\overline{P_{12}}}{\SI{d2}{\watt}}$} & \multicolumn{1}{ c }{$\frac{\overline{|\Delta P|}}{\SI{d2}{\watt}}$} & \multicolumn{1}{ c}{$\frac{\smallint\!\delta P}{\SI{d2}{\joule}}$} \\
		\noalign{\smallskip}\hline\noalign{\smallskip}
		constant & & 1. & -2.2 & 0.4 & 0.3 \\
		\noalign{\smallskip}\hline\noalign{\smallskip}
		adaptive & 1.0d-6 & 1. & -1.90 & 0.04 & 0.02 \\
		\noalign{\smallskip}\hline
	\end{tabular}
\end{table}

Another interesting observation can be made if we greatly lower the accuracy of the time integration in S$_2$ by choosing ${\Delta t}_{S_2} = \Delta t$.
As expected, the errors with constant step sizes are considerably larger yet.
Again, however, using the adaptive step size control based on the residual energy improves the situation significantly and helps keep the accuracy of the results within reasonable bounds.
As an example, consider the simulation results for the linear quarter car benchmark model as listed in Table~\ref{tab:quartercar_alt_results_lowS2}, where the residual energy is reduced by \SI{90}{\percent}.
The situation is similar with nonlinear damping forces.
This is rather beneficial for industrial applications where one does not necessarily have access to internal simulator settings in order to satisfy accuracy and stability demands on co\hyp{}simulation results.

\section{Comparison to Predictor/Corrector Method}
\label{sec:comparison_busch}

Now that we have investigated the performance and capabilities of a non-iterative energy\hyp{}conservation\hyp{}based approach to error estimation and step size control, let us in the present section deploy the quarter car benchmark model one last time to draw a comparison with the predictor/corrector method proposed by Busch \emph{et~al}.~\cite{Busch2011,Busch2012}.
To this end, we use the PI\hyp{}controller~\eqref{equ:step_control} with the error indicator~\eqref{equ:error_indicator_Busch} as outlined in Sec.~\ref{subsec:error_estimator_Busch}.
The weight between the relative and absolute errors is set to $\rho = \num{1.0d-4}$, and the tolerance $\text{TOL}$ is chosen according to the desired accuracy.
All remaining parameters are configured according to Table~\ref{tab:controller_configuration}.
Because our main focus lies on co-simulation with constant input extrapolation, we set the polynomial extrapolation order to $r = 1$.

\begin{table}[h!tb]
	\caption{%
		Linear quarter car benchmark results with predictor/corrector (pred./corr.) and residual\hyp{}energy\hyp{}based (ECCO) step size control
	}
	\label{tab:quartercar_Busch}
	\centering
	\begin{tabular}{lS[table-format=1.1d-1]S[table-format=1.1]S[table-format=1.1]S[table-format=1.1]S[table-format=1.1]}
		\hline\noalign{\smallskip}
  		\multicolumn{1}{c}{} & \multicolumn{2}{c}{Algorithm} & \multicolumn{1}{ c }{Power} & \multicolumn{2}{ c }{Error} \\
 		\multicolumn{1}{ c }{} & \multicolumn{1}{ c }{tolerance} & \multicolumn{1}{ c }{$\frac{\overline{\Delta t}}{\si{\milli\second}}$} & \multicolumn{1}{ c }{$\frac{\overline{P_{12}}}{\si{\watt}}$} & \multicolumn{1}{ c }{$\frac{\overline{|\Delta P|}}{\si{\watt}}$} & \multicolumn{1}{ c}{$\frac{\smallint\!\delta P}{\si{\joule}}$} \\
		\noalign{\smallskip}\hline\noalign{\smallskip}
		constant & & 1. & 0.4 & 1.3 & 6.4 \\
		\noalign{\smallskip}\hline\noalign{\smallskip}
		pred./corr.\ & 6.7d-1 & 1. & 0.3 & 0.7 & 2.9 \\
		\noalign{\smallskip}\hline\noalign{\smallskip}
		ECCO & 2.8d-6 & 1. & 0.0 & 0.4 & 1.6 \\
		\noalign{\smallskip}\hline
	\end{tabular}
\end{table}

\begin{table}[h!tb]
	\caption{%
		Nonlinear benchmark results with predictor/corrector (pred./corr.) and residual\hyp{}energy\hyp{}based (ECCO) step size control
	}
	\label{tab:quartercar_nonlinear_Busch}
	\centering
	\begin{tabular}{lS[table-format=1.1d-1]S[table-format=1.1]S[table-format=1.1]S[table-format=1.1]S[table-format=1.1]}
		\hline\noalign{\smallskip}
  		\multicolumn{1}{c}{} & \multicolumn{2}{c}{Algorithm} & \multicolumn{1}{ c }{Power} & \multicolumn{2}{ c }{Error} \\
 		\multicolumn{1}{ c }{} & \multicolumn{1}{ c }{tolerance} & \multicolumn{1}{ c }{$\frac{\overline{\Delta t}}{\si{\milli\second}}$} & \multicolumn{1}{ c }{$\frac{\overline{P_{12}}}{\si{\watt}}$} & \multicolumn{1}{ c }{$\frac{\overline{|\Delta P|}}{\si{\watt}}$} & \multicolumn{1}{ c}{$\frac{\smallint\!\delta P}{\si{\joule}}$} \\
		\noalign{\smallskip}\hline\noalign{\smallskip}
		constant & & 1. & 1 & 4 & 5 \\
		\noalign{\smallskip}\hline\noalign{\smallskip}
		pred./corr.\ & 2.1d0 & 1. & 0.4 & 1.9 & 3.1 \\
		\noalign{\smallskip}\hline\noalign{\smallskip}
		ECCO & 7.5d-6 & 1. & 0.0 & 1.1 & 1.6 \\
		\noalign{\smallskip}\hline
	\end{tabular}
\end{table}

\begin{table}[h!tb]
	\caption{%
		Linear benchmark results for alternative system reticulation with predictor/corrector (pred./corr.) and residual\hyp{}energy\hyp{}based (ECCO) step size control
	}
	\label{tab:quartercar_alt_Busch}
	\centering
	\begin{tabular}{lS[table-format=1.1d-1]S[table-format=1.1]S[table-format=-1.3]S[table-format=1.3]S[table-format=1.3]}
		\hline\noalign{\smallskip}
		\multicolumn{1}{c}{} & \multicolumn{2}{c}{Algorithm} & \multicolumn{1}{ c }{Power} & \multicolumn{2}{ c }{Error} \\
		\multicolumn{1}{ c }{} & \multicolumn{1}{ c }{tolerance} & \multicolumn{1}{ c }{$\frac{\overline{\Delta t}}{\si{\milli\second}}$} & \multicolumn{1}{ c }{$\frac{\overline{P_{12}}}{\SI{d2}{\watt}}$} & \multicolumn{1}{ c }{$\frac{\overline{|\Delta P|}}{\SI{d2}{\watt}}$} & \multicolumn{1}{ c}{$\frac{\smallint\!\delta P}{\SI{d2}{\joule}}$} \\
		\noalign{\smallskip}\hline\noalign{\smallskip}
		constant & & 1. & -1.92 & 0.12 & 0.23 \\
		\noalign{\smallskip}\hline\noalign{\smallskip}
		pred./corr.\ & 6.0d-1 & 1. & -1.877 & 0.013 & 0.017 \\
		\noalign{\smallskip}\hline\noalign{\smallskip}
		ECCO & 9.1d-7 & 1. & -1.879 & 0.013 & 0.016 \\
		\noalign{\smallskip}\hline
	\end{tabular}
\end{table}

\begin{table}[h!tb]
	\caption{%
		Nonlinear benchmark results for alternative system reticulation with predictor/corrector (pred./corr.) and residual\hyp{}energy\hyp{}based (ECCO) step size control
	}
	\label{tab:quartercar_alt_nonlinear_Busch}
	\centering
	\begin{tabular}{lS[table-format=1.1d-1]S[table-format=1.1]S[table-format=-1.3]S[table-format=1.3]S[table-format=1.3]}
		\hline\noalign{\smallskip}
		\multicolumn{1}{c}{} & \multicolumn{2}{c}{Algorithm} & \multicolumn{1}{ c }{Power} & \multicolumn{2}{ c }{Error} \\
		\multicolumn{1}{ c }{} & \multicolumn{1}{ c }{tolerance} & \multicolumn{1}{ c }{$\frac{\overline{\Delta t}}{\si{\milli\second}}$} & \multicolumn{1}{ c }{$\frac{\overline{P_{12}}}{\SI{d2}{\watt}}$} & \multicolumn{1}{ c }{$\frac{\overline{|\Delta P|}}{\SI{d2}{\watt}}$} & \multicolumn{1}{ c}{$\frac{\smallint\!\delta P}{\SI{d2}{\joule}}$} \\
		\noalign{\smallskip}\hline\noalign{\smallskip}
		constant & & 1. & -3.9 & 0.3 & 0.5 \\
		\noalign{\smallskip}\hline\noalign{\smallskip}
		pred./corr.\ & 6.5d0 & 1. & -3.92 & 0.18 & 0.21 \\
		\noalign{\smallskip}\hline\noalign{\smallskip}
		ECCO & 2.4d-5 & 1. & -3.77 & 0.05 & 0.05 \\
		\noalign{\smallskip}\hline
	\end{tabular}
\end{table}

The thus defined predictor/corrector method generally yields significant error reductions throughout the various benchmarks:
Errors are reduced by \SI{40}{\percent} to \SI{55}{\percent} compared to the cases with constant step sizes, see Tables~\ref{tab:quartercar_Busch} and \ref{tab:quartercar_nonlinear_Busch}.
For the alternative system reticulation discussed in Sec.~\ref{subsec:alt_reticulation}, a reduction of between \SI{40}{\percent} and \SI{93}{\percent} is achieved, see Tables~\ref{tab:quartercar_alt_Busch} and \ref{tab:quartercar_alt_nonlinear_Busch}.
In comparison, the residual\hyp{}energy\hyp{}based adaptive step size controller generally leads to a more substantial reduction in the energy errors, though both approaches achieve relatively very high accuracies for the linear benchmark model with alternative system reticulation, see Table~\ref{tab:quartercar_alt_Busch}.

%

It should be noted that the error indicator~\eqref{equ:error_indicator_Busch} is sensitive to the scale of the output values:
Recalling the corresponding discussion in Sec.~\ref{subsec:error_estimator_power}, the output values representing the spring force $F_\text{c}$~\eqref{equ:quartercar_Fc} are typically several orders of magnitude larger than the ones representing the velocities $\dot{z}_\text{c}$ and $\dot{z}_\text{w}$.
This effectively completely disregards the error contributions of the simulator outputting the velocity in Eq.~\eqref{equ:error_indicator_Busch}, and leads to an inaccurate representation of the global co\hyp{}simulation coupling error.
By using individual simulator-specific values $\text{TOL}_\alpha$ and $\rho_\alpha$, this situation can be mitigated and higher accuracies achieved.
Doing so increases the number of tunable parameters to four, however.

The particular benchmark cases at hand also reveal that the actual step size selection suggested by the predictor/corrector method as implemented according to Ref.~\onlinecite{Busch2011,Busch2012} is highly oscillatory.
This is likely due to improper tuning of the PI\hyp{}controller, and has a detrimental impact on the performance of the method for the problem set at hand.\footnote{%
	Setting $k_\text{P} = 0$ and substantially lowering $k_\text{I}$, for example, seems to be a much better choice to configure the PI-controller in this case.
}
It is also not unlikely that these issues are tied to the fact that the error contribution of one simulator is effectively disregarded.

In any case, we expect the predictor/corrector approach to yield a performance comparable to ECCO if
\romannumeral 1.) the sensitivity issue with different scales of the outputs is resolved, and
\romannumeral 2.) the PI\hyp{}controller is correctly tuned for the problem at hand.

\section{Conclusion}
\label{sec:conclusion}

When simulators are coupled via power bonds their inputs and outputs are given as so-called power variables whose product is a physical power.
Then, the flow of energy throughout a co\hyp{}simulation can be conveniently studied.
Moreover, because the subsystems are solved independently of each other between discrete communication points, energy residuals emerge.
These directly alter the total energy of the overall coupled system and distort its dynamic behavior.

In the present paper, we demonstrated that such residuals are easily computable with the simulator input and output values only.
Because they are a direct expression of the coupling errors, they constitute a novel and versatile energy\hyp{}based error estimation method.
We showed how energy residuals can be used for adaptive control of the co\hyp{}simulation step size.
The performance and applicability of this non-iterative \emph{Energy\hyp{}Conservation\hyp{}based Co\hyp{}Simulation} algorithm (ECCO) was investigated using a quarter car benchmark model, both with linear and nonlinear damping characteristics.
The proposed method ensures that approximately the correct amount of energy is exchanged between subsystems.
Consequently, significant improvements in accuracy and efficiency were demonstrated in comparison with constant co\hyp{}simulation step sizes.
Additionally, using the proposed adaptive step size control makes the accuracy of the global result less dependent on the system reticulation or the accuracies of the time integration methods inside the subsystems.

Unlike almost all other proposed co\hyp{}simulation algorithms, ECCO makes do without two major restrictions:
\romannumeral 1.) It does not require the repetition of entire co\hyp{}simulation steps.
Such rollback is often prohibited by use of commercial software tools and computationally expensive.
\romannumeral 2.) It does not require knowledge of any simulator-internal information and, consequently, supports the protection of sensitive information and intellectual property rights.
With the traction that the \emph{Functional Mock-up Interface} (FMI)~\cite{Blochwitz2011,Blochwitz2012} has been gaining over the past years~\cite{Bertsch2014,Schierz2012,Arnold2013,Viel2014,Clauss2012}, the fulfillment of such properties is a major advantage.
FMI is a tool independent standard for the exchange and the co-simulation of dynamic models that are, for most practical purposes, closed for inspection and modification.
In addition, while FMI does support the repetition of macro steps and exposing derivatives along with the outputs, such traits are still rarely found in models for co-simulation that are in use in the industry.
This makes ECCO especially attractive for industrial and engineering applications.

The same holds true for the predictor/corrector error estimation and step size control method proposed by Busch \emph{et~al}.~\cite{Busch2011,Busch2012}.
It too showed significant improvements in accuracy compared to using constant step sizes, albeit less so than ECCO.
This is likely due to its sensitivity to the scaling of the output values and its improperly tuned PI\hyp{}controller, however.
One major drawback of a predictor/corrector\hyp{}based approach to error estimation is the fact that it requires sufficiently small macro time steps to begin with.
Residual energies, on the other hand, are an exact representation of the local errors in the power outputs of two coupled simulators, irrespective of macro or micro step size.

It should be noted that using power variables to realize simulator coupling brings about three major advantages:
\romannumeral 1.) For one, power and energy are the universal currencies of physical systems, and power bonds make them directly accessible.
\romannumeral 2.) Power bonds provide a complete and universal, energy\hyp{}flow-centered connectivity between mathematical models of different engineering and physical domains.
\romannumeral 3.) Power variables typically represent higher derivatives than the coupling variables of other coupling schemes.
In general, this is a favorable trait if we keep in mind that numerical integration is much more desirable than numerical differentiation.
On the other hand, very few models and tools use power bonds to date, which may make other methods more advantageous in practice, at least for the time being.

An issue we only touched upon briefly is the stability of co\hyp{}simulations.
Due to the complexity of the topic (different solver methods of various orders, different simulator coupling schemes of various orders, the presence of direct feed-through and algebraic loops, etc.), a general approach is far from trivial and theoretical treatments lack behind practical implementations.
For further details see, for example, Refs.~\onlinecite{Busch2012,Arnold2007,Schweizer2014b,Kuebler2000,Skjong2016b}, and references therein.

\begin{acknowledgments}

This work was funded by the Research Council of Norway (grant number 225322), and the industrial partners in the ViProMa project consortium (VARD, Rolls-Royce Marine, and DNV GL).
We are grateful for their financial support.

\end{acknowledgments}

\nocite{*}    
\bibliographystyle{ieeetr}
\bibliography{energy}

\end{document}